\documentstyle[amssymb,aps]{revtex}

\begin{document}
\title{Compatibility Relations between the Reduced and Global Density Matrixes}
\author{Yong-Jian Han, Yong-Sheng Zhang, Guang-Can Guo}
\address{Key Laboratory of Quantum Information, University of Science and\\
Technology of China, Hefei 230026, China}
\maketitle

\begin{abstract}
It is a hard and important problem to find the criterion of the set of
positive-definite matrixes which can be written as reduced density operators
of a multi-partite quantum state. This problem is closely related to the
study of many-body quantum entanglement which is one of the focuses of
current quantum information theory. We give several results on the necessary
compatibility relations between a set of reduced density matrixes,
including: (i) compatibility conditions for the one-party reduced density
matrixes of any $N_{A}\times N_{B}$ dimensional bi-partite mixed quantum
state, (ii) compatibility conditions for the one-party and two-party reduced
density matrixes of any $N_{A}\times N_{B}\times N_{C}$ dimensional
tri-partite mixed quantum state, and (iii) compatibility conditions for the
one-party reduced matrixes of any $M$-partite pure quantum state with the
dimension $N^{\otimes M}$.

{\bf PACS number(s):} 03.67.-a, 03.65.Ta
\end{abstract}

\section{ Introduction}

\smallskip Any quantum states (pure or mixed) in finite dimensional Hilbert
space can be represented by positive-definite matrixes. If the Hilbert space 
$C$ has a tensor product structure $C=C_{1}\otimes C_{2}\otimes \cdots
\otimes C_{M}$, given a known global density matrix $\rho _{12\cdots M}$ in
the space $C$ for the composite system, it is straightforward to calculate
the reduced density matrixes $\rho _{\alpha }$ $\left( \alpha =1,2,\cdots
,M\right) $\ for each subsystem $\alpha $. However, the reverse of this
problem becomes much more involved. Given some positive-definite matrixes $%
\rho _{\alpha }$, it is hard to determine whether they can be written as
reductions of some global quantum state. This is what the compatibility
problem concerns about. We can list, for instance, the following
comparability problems:

\begin{itemize}
\item  Find out the criteria for the one-party density matrixes $\rho
_\alpha $ $\left( \alpha =1,2,\cdots ,M\right) $ so that they can be written
as reduced density matrixes of some global quantum state $\rho _{12\cdots M}$
in the whole Hilbert space $C$. In this case, we typically put some
restrictions on the global state $\rho _{12\cdots M}$, for instance, we may
require $\rho _{12\cdots M}$ to be pure or to have a known spectrum
(eigenvalues). Otherwise, the trivial assignment $\rho _{12\cdots M}=\rho
_1\otimes \rho _2\otimes \cdots \otimes \rho _M$ makes the above question
meaningless. Pure states actually correspond to a special class of the
known-spectrum states $\rho _{12\cdots M}$ with the eigenvalues $\lambda
_1^{\downarrow }=1$ and $\lambda _2^{\downarrow }=\lambda _3^{\downarrow
}=\cdots =0$ ($\lambda _i^{\downarrow }$ $\left( i=1,2,\cdots \right) $ are
arranged in the decreasing order)

\item  Find out the criteria for the two-party (or multi-party) density
matrixes $\rho _{\alpha \beta }$ $\left( \alpha ,\beta =1,2,\cdots ,M\right) 
$ so that they can be expressed as reduced density matrix of some global
quantum state $\rho _{12\cdots M}$. This problem is much more involved than
the first one, and also much more important for various applications for
which we will mention several examples. In this case, even without any
requirement on the global state $\rho _{12\cdots M}$, in general there is no
trivial solution to the above problem.
\end{itemize}

The investigation of the compatibility conditions between the local and the
global quantum states has some important implications: Firstly, this problem
is closely related to the study of multi-partite entanglement, which is one
of the focuses of current quantum information theory \cite
{preskill,osb,ost,lat,par,nelson}. The complexity of the compatibility
conditions really comes from the many-body entanglement inherent in the
global quantum state $\rho _{12\cdots M}$. If we restrict the global state $%
\rho _{12\cdots M}$ to be separable states, the compatibility condition
would be very simple. On the other hand, some general understanding of the
compatibility conditions will shed new light on the properties of
multi-partite entanglement. Although we have achieved remarkable
understanding of bi-partite entanglement through consideration of local
operations and classical communications, we still know little about
multi-partite entanglement. Secondly and more importantly, the understanding
of the compatibility conditions will have significant applications in
computational many-body physics \cite{colemn}. In general, the physical
interactions are sufficiently local, so the interaction energy typically
only depends on some reduced local density matrixes. If we find out some
general compatibility relations between the reduced and the global quantum
states, we could design some powerful variational approaches to solve
various many-body problems. Finally, the study of the compatibility
conditions would also help us to get a better understanding of the basic
structure of quantum mechanics and the associated Hilbert space \cite
{erhal,wootters1,wootters2}.

What do we know now about the compatibility relations between the local and
the global quantum states? There are several interesting results concerning
the compatibility conditions in some special cases. First, the well-known
GHJW theorem in quantum information theory \cite{preskill,neilson} can be
considered as a compatibility condition with the global state restricted to
be a pure bi-partite quantum state in the Hilbert space $C_{A}^{N}\otimes
C_{B}^{N}$, where $N$ denotes the dimension of each subsystem. The
compatibility criterion in this case is that the two reduced density
matrixes should have the same spectrum (eigenvalues). There are some
generalizations of this result. In particular, the compatibility criteria
between the one-party reduced density matrixes have been found in some
recent works \cite{sudbery,braviyi} if the global quantum state is
restricted to be a pure state in the Hilbert spaces $(C^{2})^{\otimes M}$, $%
C^{2}\otimes C^{2}\otimes C^{4}$, or $C^{3}\otimes C^{3}\otimes C^{3}$.

In this paper, we provide several new results concerning the compatibility
conditions in some more general cases. We derive pretty strong necessary
conditions for compatibility between the one-party reduced density matrixes
for the cases that the global quantum state is a bi-partite mixed state in
the Hilbert space $C^{N_A}\otimes C^{N_B}$ (the dimensions $N_A$ and $N_B$
are arbitrary) with a known eigenvalue spectrum $\left\{ \lambda
_i^{AB}\right\} $, or an $M$-party pure state in the Hilbert space $%
(C^N)^{\otimes M}$. Note that the previously known compatibility conditions
for the Hilbert spaces $(C^2)^{\otimes M}$ and $C^2\otimes C^2\otimes C^4$
(a pure state in $C^2\otimes C^2\otimes C^4$ corresponds to a mixed state in 
$C^2\otimes C^2$) can be considered as special cases of the results derived
here. We also consider the compatibility conditions between the two-party
density matrixes for the first time, and derive some necessary compatibility
conditions when the global state is a general tri-partite mixed state in the
Hilbert space $C^{N_A}\otimes C^{N_B}\otimes C^{N_C}$ (the dimensions $N_A$, 
$N_B$, and $N_C$ are arbitrary) with a know eigenvalue spectrum $\left\{
\lambda _i^{ABC}\right\} $. All these results are derived from a unified
mathematical method.

The paper is arranged as follows: in Sec. II, after mention of some
mathematical results which are critical for our derivation, we derive the
necessary compatibility conditions for the reduced density matrices from
mixed bi-partite quantum states; then in Sec III, we generalize this result
and derive the necessary compatibility conditions for the one-party and
two-party reduced density matrices from general mixed tri-partite quantum
states; and finally, in Sec. IV, we derive the necessary compatibility
conditions for the one-party reduced density matrices from any $M$-partite
pure quantum state with the dimension $N^{\otimes M}$.

\section{Compatibility relations of reduced density matrixes from mixed
bi-partite quantum states}

Our methods for derivation of the compatibility relations extensively use
two mathematical lemmas. First let us summarize these two lemmas:

In matrix analysis, there is an important theorem which connects the
minimization of a matrix with the matrix' eigenvalues. That is the content
of the following lemma \cite{horn}

{\bf Lemma 1}: Let $A$ denote an $n\times n$ Hermitian matrix, and $U$
denote any $n\times r$ matrix with $U^{\ast }U=I_{r}$ ($1\leq r\leqslant n$,
and $U^{\ast }$ is the adjoint of the matrix $U$)$.$ Then the minimization 
\begin{equation}
\min_{U}(tr(U^{\ast }AU))=\lambda _{1}^{\uparrow }(A)+\lambda _{2}^{\uparrow
}(A)+\cdots +\lambda _{r}^{\uparrow }(A),  \eqnum{1'}
\end{equation}
where the eigenvalues $\lambda _{1}^{\uparrow }(A),$ $\lambda _{2}^{\uparrow
}(A),\cdots ,\lambda _{r}^{\uparrow }(A)$ are arranged in the increasing
order.

In some cases, not all the column vectors of the matrix U are orthogonal to
each other. We can get another more convenient lemma.

{\bf Lemma 2}: Let $A$ denote an $n\times n$ Hermitian matrix and $U\in
M_{n\times s}$, each column of the matrix $U$ are normalized. The columns of 
$U$ can be divided into two groups, and the columns in the same group are
orthogonal each other. Suppose the linear dependent number $\varkappa
=\sum_{i,j}\left| u_i^{a*}u_j^b\right| ^2$ is an integer (where $u_i^a$ and $%
u_j^b$ are the columns in the first and the second group, respectively).
Then we have the following conclusion 
\begin{equation}
\min_U(tr(U^{*}AU))\geq \sum_{i=1}^\varkappa \lambda _i^{\uparrow
}(A)+\sum_{i=1}^{s-\varkappa }\lambda _i^{\uparrow }(A)  \eqnum{1}
\end{equation}
where $\lambda _1^{\uparrow }(A),$ $\lambda _2^{\uparrow }(A),\cdots
,\lambda _n^{\uparrow }(A)$ are the eigenvalues of the matrix $A$ and
arranged in increasing order.

Proof: Without loss of generality, we suppose that the vectors $%
u_1^a,u_2^a,\cdots ,u_l^a$ are the columns in the group $I$ and the vectors $%
u_1^b,u_2^b,\cdots ,u_{s-l}^b$ are the columns in the group $II$ . We
suppose the eigenvectors of the matrix $A$ are $v_1,v_2,\cdots ,v_n,$
corresponding to $\lambda _1^{\uparrow }(A),$ $\lambda _2^{\uparrow
}(A),\cdots ,\lambda _n^{\uparrow }(A),$ respectively. So all of the vectors
in the group $I$ and group $II$ can be expanded by the eigenvectors, that is 
\begin{eqnarray*}
u_1^a &=&\alpha _{11}v_1+\alpha _{12}v_2+\cdots +\alpha _{1n}v_n, \\
u_2^a &=&\alpha _{21}v_1+\alpha _{22}v_2+\cdots +\alpha _{2n}v_n, \\
&&\vdots \\
u_l^a &=&\alpha _{l,1}v_1+\alpha _{l,2}v_2+\cdots +\alpha _{l,n}v_n.
\end{eqnarray*}
and 
\begin{eqnarray*}
u_1^b &=&\beta _{11}v_1+\beta _{12}v_2+\cdots +\beta _{1n}v_n, \\
u_2^b &=&\beta _{21}v_1+\beta _{22}v_2+\cdots +\beta _{2n}v_n, \\
&&\vdots \\
u_{s-l}^b &=&\beta _{s-l,1}v_1+\beta _{s-l,2}v_2+\cdots +\beta _{s-l,n}v_n.
\end{eqnarray*}
Since the vectors in the same group are normalized and orthogonal to each
other, the indexes satisfy $l\leq n$ and $s-l\leq n.$ We can add some
vectors into each group to make this group a complete basis of the space. So
the coefficients must satisfy the requirements $\sum_{j=1}^l\left| \alpha
_{ji}\right| ^2\leq 1$ and $\sum_{i=1}^{s-l}\left| \beta _{ij}\right| ^2\leq
1$ . Now we can get the formula 
\begin{eqnarray*}
\min_U(tr(U^{*}AU)) &=&\min_U(\sum_{i=1}^s\sum_{j=a}^bu_i^{j*}Au_i^j) \\
&=&\min_U(\sum_{j=1}^n(\sum_{i=1}^l\left| \alpha _{ij}\right|
^2+\sum_{i=1}^{s-l}\left| \beta _{ij}\right| ^2)\lambda _j^{\uparrow }(A))
\end{eqnarray*}

To make the former function $\sum_{j=1}^n(\sum_{i=1}^l\left| \alpha
_{ij}\right| ^2+\sum_{i=1}^{s-l}\left| \beta _{ij}\right| ^2)\lambda
_j^{\uparrow }(A)$ smaller, we must make the coefficients before the smaller
eigenvalues more bigger. Since the constraints $\sum_{j=1}^l\left| \alpha
_{ji}\right| ^2\leq 1$ and $\sum_{i=1}^{s-l}\left| \beta _{ij}\right| ^2\leq
1,$ the coefficients before the eigenvalues are not more than $2$. However,
there is another constraint between the coefficients $\alpha _{ij}$ and $%
\beta _{ij},$ that is, $\varkappa =\sum_{i,j}\left| u_i^{a*}u_j^b\right| ^2$
. It means that the number of the coefficients $2$ before the eigenvalues is
not more than $\varkappa $. Under these constraints, we can get the lower
bound of this function, that is, 
\[
\min_U(tr(U^{*}AU))\geq \sum_{i=1}^\varkappa \lambda _i^{\uparrow
}(A)+\sum_{i=1}^{s-\varkappa }\lambda _i^{\uparrow }(A). 
\]
This is the end of the proof.

It is need emphasis that $s-\varkappa $ is equal to the least number of the
linear independent columns when the parameter $\varkappa $ is a integer.
When the number $\varkappa $ is not a integer, the similar results can be
found easily. We can find at the following that the integer situation is
enough for our propose.

The situation that the matrix $U$ can be divided into more than two groups
(the columns in the same group are orthogonal to each other) is more
difficult to deal with. But, for principle, we can get the similar results
by carefully calculating the linear dependent number $\varkappa $. Though
the general case is very complex, We will give a special case in the Sec IV..

Now we use these lemmas to a multi-partite density matrix to get some
relations between the reduced density matrix and the global matrix. When the
density matrix is a two particle density matrix, and each particle are
qubit, the original density matrix of the two qubit can be written as 
\[
\rho _{AB}=\left[ 
\begin{array}{cccc}
a_{00,00} & a_{00,01} & a_{00,10} & a_{00,11} \\ 
a_{01,00} & a_{01,01} & a_{01,10} & a_{01,11} \\ 
a_{10,00} & a_{10,01} & a_{10,10} & a_{10,11} \\ 
a_{11,00} & a_{11,01} & a_{11,10} & a_{11,11}
\end{array}
\right] 
\]

We can get the reduced density matrix of $\rho _A$ as 
\begin{eqnarray*}
\rho _A &=&\left[ 
\begin{array}{ll}
a_{00,00} & a_{00,10} \\ 
a_{10,00} & a_{10,10}
\end{array}
\right] +\left[ 
\begin{array}{ll}
a_{01,01} & a_{01,11} \\ 
a_{11,01} & a_{11,11}
\end{array}
\right] \\
&=&A_0+A_1
\end{eqnarray*}

For this case we have the following theorem

{\bf Theorem 1}. For the two-qubit density matrix $\rho _{AB},$ the
eigenvalues between $\rho _A,\rho _B$ and $\rho _{AB}$ have the following
relations 
\begin{equation}
\lambda _1^{\uparrow }(A)\geq \lambda _1^{\uparrow }(AB)+\lambda
_2^{\uparrow }(AB)  \eqnum{2.1}
\end{equation}

\begin{equation}
\lambda _1^{\uparrow }(B)\geq \lambda _1^{\uparrow }(AB)+\lambda
_2^{\uparrow }(AB)  \eqnum{2.2}
\end{equation}
\begin{equation}
\lambda _1^{\uparrow }(A)+\lambda _1^{\uparrow }(B)\geq 2\lambda
_1^{\uparrow }(AB)+\lambda _2^{\uparrow }(AB)+\lambda _3^{\uparrow }(AB). 
\eqnum{2.3}
\end{equation}
where $\lambda _1^{\uparrow }(A)$ and $\lambda _1^{\uparrow }(B)$ are the
smaller eigenvalue of $\rho _A$ and $\rho _B$ , respectively; $\lambda
_1^{\uparrow }(AB),$ $\lambda _2^{\uparrow }(AB),$ $\lambda _3^{\uparrow
}(AB),$ $\lambda _4^{\uparrow }(AB)$ are the eigenvalues of $\rho _{AB}$,
and they are arranged in increasing order.

Proof: With the lemma 1, we can get the smaller eigenvalue of the local
density matrix $\rho _A$ 
\begin{eqnarray*}
\lambda _1^{\uparrow }(A) &=&\min_U[tr(U^{*}\rho _AU)] \\
&=&\min_U[tr(U^{*}(A_0+A_1)U)] \\
&=&\min_U[tr(U^{*}A_0U)+tr(U^{*}A_1U)] \\
&=&\min_U[tr([u_1^{*}\text{ }u_2^{*}]\left[ 
\begin{array}{ll}
a_{00,00} & a_{00,10} \\ 
a_{10,00} & a_{10,10}
\end{array}
\right] \left[ 
\begin{array}{l}
u_1 \\ 
u_2
\end{array}
\right] )+tr([u_1^{*}\text{ }u_2^{*}]\left[ 
\begin{array}{ll}
a_{01,01} & a_{01,11} \\ 
a_{11,01} & a_{11,11}
\end{array}
\right] \left[ 
\begin{array}{l}
u_1 \\ 
u_2
\end{array}
\right] )] \\
&=&\min_U[tr([ 
\begin{array}{llll}
u_1^{*} & 0 & u_2^{*} & 0
\end{array}
]\left[ 
\begin{array}{cccc}
a_{00,00} & \$ & a_{00,10} & \$ \\ 
\$ & \$ & \$ & \$ \\ 
a_{10,00} & \$ & a_{10,10} & \$ \\ 
\$ & \$ & \$ & \$
\end{array}
\right] \left[ 
\begin{array}{l}
u_1 \\ 
0 \\ 
u_2 \\ 
0
\end{array}
\right] \\
&&+tr([ 
\begin{array}{llll}
0 & u_1^{*} & 0 & u_2^{*}
\end{array}
]\left[ 
\begin{array}{cccc}
\$ & \$ & \$ & \$ \\ 
\$ & a_{01,01} & \$ & a_{01,11} \\ 
\$ & \$ & \$ & \$ \\ 
\$ & a_{11,01} & \$ & a_{11,11}
\end{array}
\right] \left[ 
\begin{array}{l}
0 \\ 
u_1 \\ 
0 \\ 
u_2
\end{array}
\right]
\end{eqnarray*}
where $\$$ means that arbitrary number will make the equality hold, $u_i^{*}$
means the conjugate of $u_i$. We notice that the position of the elements $%
a_{ij,lm}(i,j,l,m=0,1)$ are the same as the position they are in matrix $%
\rho _{AB}.$ So we choose the proper numbers to make the middle matrix is
just equal to the density matrix $\rho _{AB}.$ Then 
\begin{eqnarray*}
\lambda _1^{\uparrow }(A) &=&\min_{U_1}[tr[ 
\begin{array}{llll}
u_1^{*} & 0 & u_2^{*} & 0
\end{array}
]\rho _{AB}\left[ 
\begin{array}{l}
u_1 \\ 
0 \\ 
u_2 \\ 
0
\end{array}
\right] +tr([ 
\begin{array}{llll}
0 & u_1^{*} & 0 & u_2^{*}
\end{array}
]\rho _{AB}\left[ 
\begin{array}{l}
0 \\ 
u_1 \\ 
0 \\ 
u_2
\end{array}
\right] )] \\
&=&\min_{U_1}[tr\left[ 
\begin{array}{llll}
u_1^{*} & 0 & u_2^{*} & 0 \\ 
0 & u_1^{*} & 0 & u_2^{*}
\end{array}
\right] \rho _{AB}\left[ 
\begin{array}{ll}
u_1 & 0 \\ 
0 & u_1 \\ 
u_2 & 0 \\ 
0 & u_2
\end{array}
\right] ] \\
&\geq &\min_U[tr\left[ 
\begin{array}{llll}
u_{11}^{*} & u_{12}^{*} & u_{13}^{*} & u_{14}^{*} \\ 
u_{21}^{*} & u_{22}^{*} & u_{23}^{*} & u_{24}^{*}
\end{array}
\right] \rho _{AB}\left[ 
\begin{array}{ll}
u_{11} & u_{21} \\ 
u_{12} & u_{22} \\ 
u_{13} & u_{23} \\ 
u_{14} & u_{24}
\end{array}
\right] ] \\
&=&\lambda _1^{\uparrow }(AB)+\lambda _2^{\uparrow }(AB),
\end{eqnarray*}
where vectors $\{u_{11},u_{12},u_{13},u_{14}\}$ and $%
\{u_{21},u_{22},u_{23},u_{24}\}$ are orthogonal..

With the same reason,we can obtain $\lambda _1^{\uparrow }(B)\geq \lambda
_1^{\uparrow }(AB)+\lambda _2^{\uparrow }(AB)$. It is more important that we
can use this method to calculate the relations between the smaller
eigenvalues of matrix $\rho _A$ and $\rho _B$, that is 
\begin{eqnarray*}
&&\lambda _1^{\uparrow }(A)+\lambda _1^{\uparrow }(B) \\
&=&\min_{U_1}[tr(U_1^{*}\rho _AU_1)]+\min_{U_2}[tr(U_2^{*}\rho _BU_2)] \\
&=&\min_{U_1}[tr\left[ 
\begin{array}{llll}
u_{11}^{*} & 0 & u_{12}^{*} & 0 \\ 
0 & u_{11}^{*} & 0 & u_{12}^{*}
\end{array}
\right] \rho _{AB}\left[ 
\begin{array}{ll}
u_{11} & 0 \\ 
0 & u_{11} \\ 
u_{12} & 0 \\ 
0 & u_{12}
\end{array}
\right] ]+\min_{U_2}[tr\left[ 
\begin{array}{llll}
u_{21}^{*} & u_{22}^{*} & 0 & 0 \\ 
0 & 0 & u_{21}^{*} & u_{22}^{*}
\end{array}
\right] \rho _{AB}\left[ 
\begin{array}{ll}
u_{21} & 0 \\ 
u_{22} & 0 \\ 
0 & u_{21} \\ 
0 & u_{22}
\end{array}
\right] ]
\end{eqnarray*}
Since the variables in $U_1$ are independent on the variables in $U_2$,we
can combine these two matrixes into one matrix $U$, that is 
\[
\lambda _1^{\uparrow }(A)+\lambda _1^{\uparrow }(B)=\min_U[tr(U^{*}\rho
_{AB}U)] 
\]
where 
\begin{equation}
U=\left[ 
\begin{array}{cccc}
u_{11} & 0 & u_{21} & 0 \\ 
0 & u_{11} & u_{22} & 0 \\ 
u_{12} & 0 & 0 & u_{21} \\ 
0 & u_{12} & 0 & u_{22}
\end{array}
\right] .  \eqnum{I}
\end{equation}
Now using the lemma $2$, we need to calculate the maximal linear dependent
number $\varkappa $ between unitary $U_1$ and $U_2.$ Here the linear
dependent number $\varkappa $ is a constant $1$. On the other hand, we can
find that this matrix at least has three linear independent columns. Then we
get the relation 
\[
\lambda _1^{\uparrow }(A)+\lambda _1^{\uparrow }(B)\geq 2\lambda
_1^{\uparrow }(AB)+\lambda _2^{\uparrow }(AB)+\lambda _3^{\uparrow }(AB). 
\]

QED

These conditions can be viewed as the necessary conditions for the problem
whether the single-qubit reduced density matrices are compatible with a two
qubit density matrix with the eigenvalues $\{\lambda _1^{\uparrow
}(AB),\lambda _2^{\uparrow }(AB),\lambda _3^{\uparrow }(AB),\lambda
_4^{\uparrow }(AB)\}$. These conditions have already received by Bravyi\cite
{braviyi} in another way. Unfortunately, these conditions are not
sufficient, and the sufficient conditions need another condition 
\[
\left| \lambda _1^{\uparrow }(A)-\lambda _1^{\uparrow }(B)\right| \leq
min\{\lambda _3^{\uparrow }(AB)-\lambda _1^{\uparrow }(AB),\lambda
_4^{\uparrow }(AB)-\lambda _2^{\uparrow }(AB)\}. 
\]
This condition can not find by our method easily. Now we turn to consider
the general case of the two particle situation.

Suppose the dimensions of particle $A$ and particle $B$ are $L$ and $N$,
respectively. Let $\{\lambda _1^{\uparrow }(A),$ $\lambda _2^{\uparrow }(A),$
$\cdots ,$ $\lambda _L^{\uparrow }(A)\},$ $\{\lambda _1^{\uparrow }(B),$ $%
\lambda _2^{\uparrow }(B),$ $\cdots ,$ $\lambda _N^{\uparrow }(B)\}$ and $%
\{\lambda _1^{\uparrow }(AB),$ $\lambda _2^{\uparrow }(AB),$ $\cdots ,$ $%
\lambda _{LN}^{\uparrow }(AB)\}$ be the eigenvalues of the density matrix $%
\rho _A,\rho _B$ and$\rho _{AB}$ , respectively, and they are arranged in
increasing order. Before giving the following theorem, we need define
majorization relation between two vectors. Let $x=\{x_1^{\uparrow
},x_2^{\uparrow },\cdots ,x_n^{\uparrow }\}$ and $y=\{y_1^{\uparrow
},y_2^{\uparrow },\cdots ,y_n^{\uparrow }\}$ are $n$-dimensional vectors and
the elements are arranged in increasing order. Then we call the vector $x$
is majorized by vector $y$\cite{olkin}, denoted by $y\succ x,$ if for each $%
k $ $(k=1,2,\cdots ,n)$ the following inequality hold 
\[
\sum_{i=1}^kx_i^{\uparrow }\geq \sum_{i=1}^ky_i^{\uparrow } 
\]
and the equality hold when $k=n$. The majorization relation have already
been extensively used in quantum information\cite{nielsen}. If we define the
vector $\lambda _A=\{\lambda _1^{\uparrow }(A),$ $\lambda _2^{\uparrow }(A),$
$\cdots ,$ $\lambda _L^{\uparrow }(A)\},$ $\lambda _B=\{\lambda _1^{\uparrow
}(B),$ $\lambda _2^{\uparrow }(B),$ $\cdots ,$ $\lambda _N^{\uparrow }(B)\}$%
, $\lambda _{AB}^A=\{\sum_{j=1}^N\lambda _j^{\uparrow }(AB),$ $%
\sum_{j=N+1}^{2N}\lambda _j^{\uparrow }(AB),$ $\cdots ,$ $%
\sum_{j=(L-1)N+1}^{LN}\lambda _j^{\uparrow }(AB)$ $\}$ and $\lambda
_{AB}^B=\{\sum_{j=1}^L\lambda _j^{\uparrow }(AB),$ $\sum_{j=L+1}^{2L}\lambda
_j^{\uparrow }(AB),$ $\cdots ,\sum_{j=(N-1)L+1}^{NL}\lambda _j^{\uparrow
}(AB)\}.$ Using these definition, we can get the theorem for the general
bi-partite case as the following

{\bf Theorem 2}. As the note before, we get the following relations between
the eigenvalues of $\rho _A,\rho _B$ and $\rho _{AB}$, 
\begin{equation}
\lambda _{AB}^A\succ \lambda _A;  \eqnum{3.1}
\end{equation}
\begin{equation}
\lambda _{AB}^B\succ \lambda _B;  \eqnum{3.2}
\end{equation}
\begin{equation}
\sum_{i=1}^k\lambda _i^{\uparrow }(A)+\sum_{j=1}^l\lambda _j^{\uparrow
}(B)\geq \sum_{i=1}^{kN+lL-kl}\lambda _i^{\uparrow
}(AB)+\sum_{j=1}^{kl}\lambda _j^{\uparrow }(AB),k=1,2,\cdots
,L-1,l=1,2,\cdots ,N-1.  \eqnum{3.3}
\end{equation}

Before giving the proof, We need some discussions about these conditions.
The majorization relations (3.1), (3.2) between the one-party reduced
density matrix and the bi-partite density matrix are not just hold for this
special case. We can see in the following, the majorization is a universal
relations between the eigenvalues of reduced density matrices and the
multi-partite density matrix. This can be viewed as one of the reasons why
the majorization relations play an important role in the quantum
information. The relations (3.3) tell us that some equalities in the former
two relations can not be hold at the same time except for some special
situations. This fact can be viewed as the correlation between the different
reduced density matrices. When $L=N=2$, this theorem is reduced to the
theorem $1$.

{\it Proof }Using the similar method as the qubit case, we can get 
\begin{eqnarray*}
\sum_{i=1}^k\lambda _i^{\uparrow }(A) &=&\min_U[tr(U^{*}\rho _AU)] \\
&=&\min_U[tr(U^{*}A_0U)+tr(U^{*}A_1U)+\cdots +tr(U^{*}A_NU)] \\
&=&\min_{U_1}[tr(U_1^{*}\rho _{AB}U_1)]
\end{eqnarray*}

where the unitary matrix $U\in M_{L\times k}$, $U^{*}U=I\in M_k$ and $A_i$
is equal to $\left\langle 0\right| \rho _{AB}\left| 0\right\rangle .$ Using
the same method which used to construct the matrix ($I$), we can get a
unitary matrix $U_1\in M_{LN\times kN},$ we divide this matrix into $k$
blocks, and each block is a $LN\times N$ matrix which has the following form 
\begin{equation}
\left[ 
\begin{array}{cccc}
u_{1i} &  &  &  \\ 
& u_{1i} &  &  \\ 
&  & \ddots &  \\ 
&  &  & u_{1i} \\ 
u_{N+1,i} &  &  &  \\ 
& u_{N+1,i} &  &  \\ 
&  & \ddots &  \\ 
&  &  & u_{N+1,i} \\ 
u_{(L-1)N+1,i} &  &  &  \\ 
& u_{(L-1)N+1,i} &  &  \\ 
&  & \ddots &  \\ 
&  &  & u_{(L-1)N+1,i}
\end{array}
\right] _{LN\times N},  \eqnum{II}
\end{equation}
where only the elements $(pN+q,q)(p=0,1,\cdots ,L-1;q=1,2,\cdots ,N)$ in
this block are nonzero and $i$ denotes the $i$th block. The first columns in
the different blocks are orthogonal and normalized. So all of the columns in
the matrix $U_1$ are orthogonal and normalized, that is 
\[
\sum_{i=1}^k\lambda _i^{\uparrow }(A)\geq \sum_{i=1}^{kN}\lambda
_i^{\uparrow }(AB). 
\]
As the same reason, 
\begin{eqnarray*}
\sum_{i=1}^l\lambda _i^{\uparrow }(B) &=&\min_U[tr(U^{*}\rho _BU)] \\
&=&\min_U[tr(U^{*}B_0U)+tr(U^{*}B_1U)+\cdots +tr(U^{*}B_LU)] \\
&=&\min_{U_2}[tr(U_2^{*}\rho _{AB}U_2)]
\end{eqnarray*}
where the unitary matrix $U_2\in M_{LN\times Ll}.$ We divide the matrix $U_2$
into $l$ blocks, and each block is a $LN\times L$ matrix which has the
following form 
\begin{equation}
\left[ 
\begin{array}{cccc}
u_{1i} &  &  &  \\ 
u_{2i} &  &  &  \\ 
\vdots &  &  &  \\ 
u_{Ni} &  &  &  \\ 
& u_{1i} &  &  \\ 
& u_{2i} &  &  \\ 
& \vdots &  &  \\ 
& u_{Ni} &  &  \\ 
&  & \ddots &  \\ 
&  &  & u_{1i} \\ 
&  &  & u_{2i} \\ 
&  &  & \vdots \\ 
&  &  & u_{Ni}
\end{array}
\right] _{LN\times L},  \eqnum{III}
\end{equation}
where only the elements $((p-1)N+q,p)(p=1,\cdots ,L;q=1,2,\cdots ,N)$ in the 
$i$th block are nonzero. The first columns in the different blocks are also
orthogonal and normalized. So we can get 
\[
\sum_{i=1}^l\lambda _i^{\uparrow }(B)\geq \sum_{i=1}^{lL}\lambda
_i^{\uparrow }(AB). 
\]
We also need to find the relations between the eigenvalues of $\rho _A$ and $%
\rho _B$%
\[
\sum_{i=1}^k\lambda _i^{\uparrow }(A)+\sum_{j=1}^l\lambda _j^{\uparrow
}(B)=\min_{U_1}[tr(U_1^{*}\rho _AU_1)]+\min_{U_2}[tr(U_2^{*}\rho _BU_2)]. 
\]
As the lemma $2$, we need to calculate the linear dependent number between
matrix $U_1$ and $U_2.$ Here it is easy to get the linear dependent number $%
\varkappa $ , where $\varkappa =$ $\sum_{i=1}^l\sum_{j=1}^k\left|
\left\langle V_i^2\right. \left| V_j^1\right\rangle \right| ^2$( $V_i^2$ is
the $i$th column in the unitary matrix $U_1$ and $V_j^1$ is the $j$th column
in the unitary matrix $U_2$) is equal to $kl.$ So we get the result 
\[
\sum_{i=1}^k\lambda _i^{\uparrow }(A)+\sum_{j=1}^l\lambda _j^{\uparrow
}(B)\geq \sum_{i=1}^{kN+lL-kl}\lambda _i^{\uparrow
}(AB)+\sum_{j=1}^{kl}\lambda _j^{\uparrow }(AB). 
\]
QED

We can also find that there are at least $kN+lL-kl$ linear independent
columns in matrix $U_1$ and $U_2$. For convenience, we let $k\geq l,$ we
find in the following that at most $kl$ columns in unitary $U_1$ are linear
dependent on the columns in $U_2$. At first, we point out that in each block
of unitary $U_1$ at most $k$ columns are linear dependent on the columns in
unitary $U_2.$ Suppose the $j$th column of the $i$th block in the unitary $%
U_1$ is linear dependent on the columns in unitary $U_2$, then only the
first columns of the block in unitary $U_2$ contribute to the first $N$
elements of the unitary $U_1.$ So we can get 
\[
\alpha _{j1}V_1+\alpha _{j2}V_2+\cdots +\alpha _{jl}V_l=W_j. 
\]
where the vector $V_i$ is the first column of the $i$th block in the unitary 
$U_2$ and the vector $W_j$ is a $LN$ vector $\{0,\cdots ,0,u_{ij},0,\cdots
,0\}^T(u_{ij}$ is the $j$th element in this vector and is equal to the first
nonzero element of the $j$th column of $i$th block and $T$ means transpose).
Since $V_i(i=1,2,\cdots ,l)$ are orthogonal each other for different $i$,
and they span a $l$-dimensional space. Then at most $l$ orthogonal $W_i$ can
be get from $V_i(i=1,2,\cdots ,l)$ by liner combination. That is, at most $l$
columns in the $i$th block of unitary $U_1$ are linear dependent on the
columns in the unitary $U_2.$ Since there are $k$ blocks, then there at most 
$kl$ columns in the unitary $U_1$ are linear dependent on the columns in the
unitary $U_2.$ We will show that this situation can be reached by letting
the element $u_{(i-1)N+1,i}$ in the $i$th block of the unitary $U_1$ be
equal to $1$ and the other elements zeros. So we can also get the conditions 
\[
\sum_{i=1}^k\lambda _i^{\uparrow }(B)+\sum_{j=1}^l\lambda _j^{\uparrow
}(B)\geq \sum_{i=1}^{kN+lL-kl}\lambda _i^{\uparrow
}(AB)+\sum_{j=1}^{kl}\lambda _j^{\uparrow }(AB). 
\]

These conditions also can be viewed as the necessary conditions to the
problem whether the one-party reduced density matrices are compatible with
the multi-partite density matrix. The method we used to find these
conditions is very simple, and it is dependent on neither the number of the
particles nor the dimensions of the particles. During the calculation, the
most important thing is to get the maximal linear dependent number $%
\varkappa $ between the different unitary matrices.

\section{ Compatibility relations between one-party and two-party density
matrixes from tri-partite mixed quantum states}

There is a density matrix $\rho _{ABC\text{,}}$ where the particle $A,B$ and 
$C$ are in $L-$dimension, $M-$dimension and $N-$dimension Hilbert space,
respectively. Let $\{\lambda _1^{\uparrow }(AB),$ $\lambda _2^{\uparrow
}(AB),$ $\cdots ,$ $\lambda _{LM}^{\uparrow }(AB)\},$ $\{\lambda
_1^{\uparrow }(BC),$ $\lambda _2^{\uparrow }(BC),$ $\cdots ,$ $\lambda
_{MN}^{\uparrow }(BC)\},$ $\{\lambda _1^{\uparrow }(B),$ $\lambda
_2^{\uparrow }(B),$ $\cdots ,$ $\lambda _M^{\uparrow }(B)\}$ and $\{\lambda
_1^{\uparrow }(ABC),$ $\lambda _2^{\uparrow }(ABC),$ $\cdots ,$ $\lambda
_{LMN}^{\uparrow }(ABC)\}$ be the eigenvalues of the density matrix $\rho
_{AB},\rho _{BC},\rho _B$ and $\rho _{ABC}$, respectively, and they are
arranged in increasing order. For convenience, We define the following
vectors $\lambda _{AB}=\{\lambda _1^{\uparrow }(AB),$ $\lambda _2^{\uparrow
}(AB),$ $\cdots ,$ $\lambda _{LM}^{\uparrow }(AB)\},$ $\lambda
_{BC}=\{\lambda _1^{\uparrow }(BC),$ $\lambda _2^{\uparrow }(BC),$ $\cdots
,\lambda _{MN}^{\uparrow }(BC)\},$ $\lambda _B=\{\lambda _1^{\uparrow }(B),$ 
$\lambda _2^{\uparrow }(B),$ $\cdots ,$ $\lambda _M^{\uparrow }(B)\},$ $%
\lambda _{ABC}=\{\lambda _1^{\uparrow }(ABC),$ $\lambda _2^{\uparrow }(ABC),$
$\cdots ,$ $\lambda _{LMN}^{\uparrow }(ABC)\}$ and $\lambda
_{AB}^B=\{\sum_{i=1}^L\lambda _i^{\uparrow }(AB),$ $\sum_{i=L+1}^{2L}\lambda
_i^{\uparrow }(AB),$ $\cdots ,$ $\sum_{i=L(M-1)+1}^{LM}\lambda _i^{\uparrow
}(AB)\},$ $\lambda _{BC}^B=\{\sum_{i=1}^N\lambda _i^{\uparrow }(BC),$ $%
\sum_{i=N+1}^{2N}\lambda _i^{\uparrow }(BC),$ $\cdots ,$ $%
\sum_{i=(M-1)N+1}^{MN}\lambda _i^{\uparrow }(BC)\},$ $\lambda
_{ABC}^{AB}=\{\sum_{j=1}^N\lambda _j^{\uparrow }(ABC),$ $\sum_{j=N+1}^{2N}%
\lambda _j^{\uparrow }(ABC),$ $\cdots ,$ $\sum_{j=(LM-1)N+1}^{LMN}\lambda
_j^{\uparrow }(ABC)$ $\},$ $\lambda _{ABC}^{BC}=\{\sum_{j=1}^L\lambda
_j^{\uparrow }(ABC),$ $\sum_{j=L+1}^{2L}\lambda _j^{\uparrow }(ABC),$ $%
\cdots ,$ $\sum_{j=(MN-1)L+1}^{LMN}\lambda _j^{\uparrow }(ABC)$ $\}.$ For
this situation, we can get the following theorem.

{\bf Theorem 3}. Using the notes defined before, we can get the relations
between the eigenvalues of $\rho _{BC}$ , $\rho _{AB\text{,}}$ $\rho _B$ and 
$\rho _{ABC}$ as 
\begin{equation}
\lambda _{ABC}^{AB}\succ \lambda _{AB}  \eqnum{4.1}
\end{equation}
\begin{equation}
\lambda _{ABC}^{BC}\succ \lambda _{BC}  \eqnum{4.2}
\end{equation}
\begin{eqnarray}
\lambda _{BC}^B &\succ &\lambda _B  \eqnum{4.3.1} \\
\lambda _{AB}^B &\succ &\lambda _B  \eqnum{4.3.2}
\end{eqnarray}
\begin{equation}
\sum_{i=1}^{\mu L+r}\lambda _i^{\uparrow }(AB)+\sum_{j=1}^{\mu N+s}\lambda
_j^{\uparrow }(BC)\geq \sum_{k=1}^{\mu LN+Nr+Ls-rs}\lambda _k^{\uparrow
}(ABC)+\sum_{l=1}^{\mu LN+rs}\lambda _l^{\uparrow }(ABC),\mu <M,0\leq
r<L,0\leq s<N.  \eqnum{4.4}
\end{equation}

The relation (4.1), (4.2), (4.3) make sure the universality of the
majorization relations between the eigenvalues of the reduced density matrix
and the multi-partite density matrix. The conditions (4.1), (4.2) and (4.3)
can obtain as the same as in the bi-partite case easily. The relations (4.4)
are new relations, and they include more informations than the relations
(3.3). We can see in the following that we can get some further results of
this relations. We only need proof the conditions of (4.4).

{\bf Proof}: Using the method used in the bi-partite case 
\[
\sum_{i=1}^{R=\mu L+r}\lambda _i^{\uparrow }(AB)+\sum_{j=1}^{S=\mu
N+s}\lambda _j^{\uparrow }(BC)=\min_{U_1}[tr(U_1^{*}\rho
_{ABC}U_1)]+\min_{U_2}[tr(U_2^{*}\rho _{ABC}U_2)]. 
\]
where $U_1$ is a $LMN\times RN$ matrix, every $N$ columns can be viewed as a
block. So matrix $U_1$ is divided into $R$ blocks. The $i$th block has the
form 
\begin{equation}
\left[ 
\begin{array}{cccc}
u_{1i} &  &  &  \\ 
& u_{1i} &  &  \\ 
&  & \ddots &  \\ 
&  &  & u_{1i} \\ 
u_{N+1,i} &  &  &  \\ 
& u_{N+1,i} &  &  \\ 
&  & \ddots &  \\ 
&  &  & u_{N+1,i} \\ 
&  &  &  \\ 
u_{(LM-1)N+1,i} &  &  &  \\ 
& u_{(LM-1)N+1,i} &  &  \\ 
&  & \ddots &  \\ 
&  &  & u_{(LM-1)N+1,i}
\end{array}
\right]  \eqnum{IV}
\end{equation}
where only the elements $((k-1)N+j,$ $j)$ $(k=1,$ $2,\cdots ,LM;$ $j=1,$ $%
2,\cdots ,$ $N)$ are nonzero in the $i$th block, the first columns in the
different blocks are orthogonal and normalized. The form of this matrix is
the same as the matrix $U_1$ used in the general bi-partite case. But
together with the following unitary matrix $U_2$, we can find that this
situation is different from the situation in the bi-partite case. In the
bi-partite case, there is only one position that both the columns in the
blocks of $U_2$ and the columns in $U_1$ are nonzero. But in this case, it
is not. This can substantially effect the linear dependent number $\varkappa 
$ between the unitary matrixes $U_1$ and $U_2.$

The unitary matrix $U_2\in M_{LMN\times LS},$ we divide this matrix into $S$
blocks, and each block is a $LMN\times L$ matrix which has the following
form 
\begin{equation}
\left[ 
\begin{array}{cccc}
u_{1,i} &  &  &  \\ 
u_{2,i} &  &  &  \\ 
\vdots &  &  &  \\ 
u_{MN,i} &  &  &  \\ 
& u_{1,i} &  &  \\ 
& u_{2,i} &  &  \\ 
& \vdots &  &  \\ 
& u_{MN,i} &  &  \\ 
&  & \ddots &  \\ 
&  &  & u_{1,i} \\ 
&  &  & u_{2,i} \\ 
&  &  & \vdots \\ 
&  &  & u_{MN,i}
\end{array}
\right]  \eqnum{V}
\end{equation}

where only the elements $((p-1)N+q,p)(p=1,\cdots ,L;q=1,2,\cdots ,MN)$ in
the $i$th block are nonzero and the first columns in the different blocks
are also orthogonal and normalized. Now we calculate the maximal linear
dependent number $\varkappa $ between $U_1$ and $U_2.$ We need to point out
two facts about these two unitary matrices.

The first, If $R<L$ and $S<N,$ we can find at most $RS$ columns in the
unitary matrix $U_2$ are linear dependent on the columns in the unitary
matrix $U_1$ as the same discussion in the bi-partite case. The second, when 
$R=\mu L$ and $S=\mu N$, all of the columns in the unitary matrix $U_2$ can
be linear dependent on the columns in the unitary matrix $U_1$. This can be
reached by letting the elements $((j-1)MN+iN+1,1)$ in the $[(i-1)L+j]$th $%
(i=0,1,2,\cdots ,\mu -1;j=1,2,\cdots ,L)$ block of unitary $U_1$ be $1$ and
the other elements zeroes; at same time letting the elements $(k,(i-1)N+j)$ $%
(k=(i-1)N+1,(i-1)N+2,\cdots iN-1)$ in the $[(i-1)N+j]$th $(i=1,2,\cdots ,\mu
;j=1,2,\cdots ,N)$ be nonzero and the others zeroes. Using these two facts,
when $R=\mu L+r,S=\mu N+s,$ there are $\mu LN$ columns in $U_2$ linear
dependent on $\mu LN$ columns in $U_1$. Since the columns in the same
unitary matrix are orthogonal to each other, we need only consider the rest
columns in the unitary matrix $U_1$ and $U_2.$ So at most $\mu LN+rs$
columns in the unitary $U_2$ are linear dependent on the columns in the
unitary matrix $U_1.$ QED

When some eigenvalues of the density matrices are zeros, we can get some
stronger relations between the reduced density matrix and the global density
matrix.

{\bf Theorem 4}: Suppose $rank(\rho _{ABC})=LMN-Ls,$ $rank(\rho _{BC})=MN-s,$
$rank(\rho _{AB})=LM-r$ and $rank(\rho _B)=M-t,$ if $r$ and $s$ satisfy the
condition $Nr\leq Ls,$then 
\begin{equation}
t\leq [\frac{r-1}L]+1  \eqnum{5}
\end{equation}
is hold, where $[x]$ is the maximum integer which is smaller than $x$.

The proof of this theorem is a little technical. It is well known that the $%
rank$ of a $N\times N$ Hermitian matrix is equal to $N-k$ (where $k$ is the
number of the zero eigenvalues of the matrix). So we need only consider the
number of the zero eigenvalues of matrix $\rho _B.$ We construct some
orthogonal vectors which corresponding to the zero eigenvalues from some
known matrixes. In the following proof, at first, we point out that the
columns in the matrix $A$ (defined in the following) are linear combinations
of the columns in matrix $B$ (defined in the following). Then we use this
fact to construct $[\frac{r-1}L]+1$ orthogonal vectors which corresponding
to zero eigenvalues of matrix $\rho _B.$

{\it Proof }To proof this theorem, we need to define the matrix $A,B$ and $C$
for density matrix $\rho _{AB},\rho _{BC}$ and $\rho _B$, respectively$.$
The form of the matrix $A$ is the same as the matrix $U_1$ used in the proof
of the theorem 3 and made the same division. The matrix $B$ is the same as
the matrix $U_2$ in the proof of the theorem 3. but we do a different
division. We viewed each $s$ columns as a block. So the matrix $B$ is
divided into $L$ blocks. Each block has the form as a $LMN\times s$ matrix.
In the $(i+1)$th block, only the elements $(iMN+j,k)(j=1,2,\cdots ,MN;$ $%
k=1,2,\cdots ,s)$ are nonzero. We can note that the nonzero elements are not
dependent on the index $i$, that is, the element ($iMN+j,k)=v_{jk}.$ The
form of the $(i+1)$th block is

\begin{equation}
\left[ 
\begin{array}{cccc}
&  &  &  \\ 
&  &  &  \\ 
&  &  &  \\ 
v_{iMN+1,1} & v_{iMN+1,2} & \cdots & v_{iMN+1,s} \\ 
v_{iMN+2,1} & v_{iMN+2,2} & \cdots & v_{iMN+2,s} \\ 
\vdots & \vdots & \cdots & \vdots \\ 
v_{iMN+MN,1} & v_{iMN+MN,2} & \cdots & v_{iMN+MN,s} \\ 
&  &  &  \\ 
&  &  &  \\ 
&  &  & 
\end{array}
\right]  \eqnum{VI}
\end{equation}
The columns in the same block are normalized and orthogonal to each other.

The matrix $C$ can be divided into many blocks, each block is a $LMN\times
LN $ matrix, further more, we can divide each block into $L$ sub-blocks. The 
$(i+1)$th sub-block has the following form 
\begin{equation}
\left[ 
\begin{array}{cccc}
&  &  &  \\ 
w_{iMN,1} &  &  &  \\ 
& w_{iMN,1} &  &  \\ 
&  & \ddots &  \\ 
w_{iMN+N,1} &  &  & w_{jMN,1} \\ 
& w_{iMN+N,1} &  &  \\ 
&  &  &  \\ 
w_{iMN+(M-1)N,1} &  &  & w_{iMN+N,1} \\ 
& w_{iMN+(M-1)N,1} &  &  \\ 
&  & \ddots &  \\ 
&  &  & w_{iMN+(M-1)N,1} \\ 
&  &  & 
\end{array}
\right] _{LMN\times N}  \eqnum{VII}
\end{equation}
In this sub-block, only the elements $(iMN+kN+j,j)$ $(k=0,1,\cdots
,M-1;j=0,1,\cdots ,N)$ are nonzero and they are not dependent on $j.$ For
each block, the value of the nonzero element are not dependent on the index $%
i.$ The first columns of the different blocks are orthogonal and normalized.
Using the lemma 1, we can get the relations of the eigenvalues 
\[
\sum_{j=1}^r\lambda _j^{\uparrow }(AB)=\min_Atr[A^{*}\rho _{ABC}A], 
\]
\[
\sum_{k=1}^s\lambda _k^{\uparrow }(BC)=\min_Btr[B^{*}\rho _{ABC}B], 
\]
\[
\sum_{j=1}^t\lambda _j^{\uparrow }(B)=\min_Atr[C^{*}\rho _{ABC}C]. 
\]

Since $\sum_{k=1}^s\lambda _k^{\uparrow }(BC)=\sum_{k=1}^{Ls}\lambda
_k^{\uparrow }(ABC)=0,$ there is a unitary transformation between the $Ls$
minimal eigenvectors and the columns of the matrix $B$. Because of $%
\sum_{j=1}^r\lambda _j^{\uparrow }(AB)=0$ and the density matrix $\rho
_{ABC} $ has only the $Ls$ zero eigenvalues, we can see that the columns in
matrix $A$ must be a linear combination of the $Ls$ minimal eigenvectors. So
all of the columns in the matrix $A$ must be a linear combinations of the
columns of the matrix $B$. For convenience, we divide each block of the
matrix $A$ into $L$ sub-blocks, each sub-block is a $MN\times N$ matrix, the 
$(i+1)$th sub-block has the following form 
\begin{equation}
\left[ 
\begin{array}{cccc}
u_{iMN,1} &  &  &  \\ 
& u_{iMN,1} &  &  \\ 
&  & \ddots &  \\ 
u_{iMN+N,1} &  &  & u_{iMN,1} \\ 
& u_{iMN+N,1} &  &  \\ 
&  & \ddots &  \\ 
u_{iMN+(M-1)N,1} &  &  & u_{iMN+N,1} \\ 
& u_{iMN+(M-1)N,1} &  &  \\ 
&  & \ddots &  \\ 
&  &  & u_{iMN+(M-1)N,1}
\end{array}
\right] _{MN\times N}.  \eqnum{VIII}
\end{equation}

Now we can find that each columns in the $(i+1)$th sub-block of the matrix $%
A $ must be the linear combination of the columns of the $(i+1)$th block of
the matrix $B$. And we take out the first columns of all of the sub-block
and to find how many columns are linear independent. As the following proof,
there are at least $[\frac{r-1}L]+1$ columns are linear independent. At
first, we denote the column of the $(i+1)$th sub-block in the $(j+1)$th
block as $V_{ij}$. We write all these vectors in a matrix form as 
\begin{equation}
\left[ 
\begin{array}{cccc}
V_{01} & V_{02} & \cdots & V_{0r} \\ 
V_{11} & V_{12} & \cdots & V_{1r} \\ 
\vdots & \ddots & \cdots & \vdots \\ 
V_{(L-2),1} & \cdots & \cdots & V_{(L-2),r} \\ 
V_{(L-1),1} & V_{(L-1),2} & \cdots & V_{(L-1),r}
\end{array}
\right] _{L\times r}.  \eqnum{IX}
\end{equation}
The different columns in this matrix are orthogonal and normalized. At
first, we consider the columns. There are at least one linear independent
vector in each column, that is, every elements in the same columns are equal
to the same vector multiply by a scalar. So we can rewrite the matrix (IX)
as the following matrix 
\begin{equation}
\left[ 
\begin{array}{cccc}
\alpha _{1,1}V_1 & \alpha _{2,1}V_2 & \cdots & \alpha _{r,1}V_r \\ 
\alpha _{1,2}V_1 & \alpha _{2,2}V_2 & \cdots & \alpha _{r,2}V_r \\ 
\vdots & \ddots & \cdots & \vdots \\ 
\alpha _{1,L-1}V_1 & \cdots & \cdots & \alpha _{r,L-1}V_r \\ 
\alpha _{1,L}V_1 & \alpha _{2,L}V_2 & \cdots & \alpha _{r,L}V_r
\end{array}
\right] _{L\times r},  \eqnum{X}
\end{equation}
where $\alpha _{i,j}(i=1,2,\cdots r,j=1,2,\cdots ,L)$ are complex number.
Now we use the orthogonal conditions that the different columns are
orthogonal each other. That is, 
\[
\left\langle \alpha _i\right. \left| \alpha _j\right\rangle \left\langle
V_i\right. \left| V_j\right\rangle =0\text{ }i,j=1,2,\cdots ,r 
\]
where the vector $\left| \alpha _j\right\rangle $ is $(\alpha _{1,1},\alpha
_{1,2},\cdots ,\alpha _{1,L})^T$. From this equations we know that if the
vectors $V_i$ and $V_j$ are linear dependent, then the vectors $\left|
\alpha _i\right\rangle $ and $\left| \alpha _j\right\rangle $ must be
orthogonal. Since the vector $\left| \alpha _j\right\rangle $ is in $L$
Dimension space, there are at most $L$ $V_i$ are linear dependent to the
same vector $v_p$ and the vectors $\left| \alpha _j\right\rangle $ are
orthogonal each other. So there are at least $[\frac{r-1}L]+1$ linear
independent $v_p$. Using the Schmidt method, we can get $[\frac{r-1}L]+1$
orthogonal and normalized vectors $w_i(i=1,2,\cdots ,L)$. All this vectors
can be expressed as the linear combination of the columns of the matrix $B$.

If we let the first column of the $(i+1)$th sub-block of the matrix $C$ be
equal to one of the vectors $w_i$. Then the columns of the $(i+1)$th
sub-block also can be expressed as the linear combinations of the columns of
the $(i+1)$th block of the matrix $B$. Since the nonzero elements of the
matrix $B$ and $C$ are not dependent on the index of $i$, the columns in the
block which the sub-block belongs to can be expressed as the linear
combination of the columns of the matrix $B$. So the matrix $\rho _B$ must
have at least $[\frac{r-1}L]+1$ zero eigenvalues.

QED

From the proof of this theorem, we can find that the most important fact is
that the columns in the matrix $A$ must be the linear combinations of the
columns of the matrix $B$. Since the symmetry of the particles, exchange the
role of the density matrix of $\rho _{AB}$ and $\rho _{BC},$ the former
theorem is hold too. This theorem is the character of the multi-partite
density matrix, the bi-partite situation has no theorem similar as this
theorem. This can be viewed as a new correlation between the particles. This
theorem is the further results of the relations (4.4). The parameters $r$
and $s$ must satisfy the constrain of the inequality(4.4).

\section{Compatibility relations between one-party density matrixes from
N-partite pure quantum states}

When using our method to the multi-partite density matrix, we can get much
more complicated relations between the reduced density matrices and the
multi-partite density matrix. The majorization relations between them are
hold and can be gotten easily. But it is very difficult to find the
relations which are similar as the relations (3.3) and (4.4), since it is
not easy to find the maximal linear dependent number. There is a special
case where the $N$-partite density matrix is a pure state and all of the
particles are in a $M$ dimension Hilbert space, we can get some simple
formulas, that is,

{\bf Theorem 6}: For a $N$-partite pure state, if every particle are in the $%
M$-dimensional Hilbert space, the eigenvalues of the one-party reduced
density matrices satisfy the following relations 
\begin{equation}
\sum_{j=1,j\neq k,l}^N\sum_{i=1}^{M-1}\lambda _i^{\uparrow
}(j)+\sum_{i=1}^p\lambda _i^{\uparrow }(k)\geq \sum_{i=1}^p\lambda
_i^{\uparrow }(l),\text{ }p=1,2,\cdots ,M-1,k\neq l=1,2,\cdots ,N,  \eqnum{6}
\end{equation}
where $\lambda _1^{\uparrow }(j),\lambda _2^{\uparrow }(j),\cdots ,\lambda
_{M-1}^{\uparrow }(j)$ are the eigenvalues of the partial density matrix $%
\rho _j$ and they are arranged in increasing order.

If $N=2$, we can get the necessary and sufficient conditions for the single
particle partial density matrices compatible with the bi-partite pure state.
When $M=2$, it can also give the necessary and sufficient compatibility
conditions between the set of one-party reduced density matrices and the $N$%
-partite density matrix\cite{sudbery}.

{\it Proof }We consider the $(N-1)$-partite density matrix $\rho _{123\cdots
N-1}$. From this density matrix, we can get the one-party reduced density
matrices $\rho _i(i=1,2,\cdots ,N-1)$. Now we use the method before to get 
\[
\sum_{j=1}^{N-2}\sum_{i=1}^{M-1}\lambda _i^{\uparrow
}(j)+\sum_{i=1}^p\lambda _i^{\uparrow
}(N-1)=\sum_{i=1}^{N-2}\min_{U_i}[tr(U_i^{*}\rho _{12\cdots
N-1}U_i)]+\min_{U_{N-1}}[tr(U_{N-1}^{*}\rho _{12\cdots N-1}U_{N-1})] 
\]
where the unitary matrix $U_i$ $(i=1,2,\cdots ,N-2)$ can be divided into $%
M-1 $ blocks, each block is a $M^{N-1}\times M^{N-2}$ matrix. The unitary
matrix $U_{N-1}$ can be divided into $p$ blocks, each block is also a $%
M^{N-1}\times M^{N-2}$ matrix. The form of the block (the positions of the
nonzero elements in the block) is independent of the block in the same
unitary matrix $U_i$. The firsts column in different blocks of the same
unitary matrix are orthogonal and normalized. Further more, we can divide
the blocks into some sub-blocks. For the block in the unitary matrix $U_i$,
we can divide it into $M^{i-1}$ sub-blocks, the form of the $j$th sub-block
of the $k$th block in $U_i$ is the following 
\begin{equation}
\left[ 
\begin{array}{cccc}
&  &  &  \\ 
&  &  &  \\ 
u_{(j-1)M^{N-i}+1,k} &  &  &  \\ 
& u_{(j-1)M^{N-i}+1,k} &  &  \\ 
&  & \ddots &  \\ 
&  &  & u_{(j-1)M^{N-i}+1,k} \\ 
u_{(j-1)M^{N-i}+M^{N-i-1}+1,k} &  &  &  \\ 
& u_{(j-1)M^{N-i}+M^{N-i-1}+1,k} &  &  \\ 
&  & \ddots &  \\ 
&  &  & u_{(j-1)M^{N-i}+M^{N-i-1}+1,k} \\ 
u_{jM^{N-i}-M^{N-i-1}+1,k} &  &  &  \\ 
& u_{jM^{N-i}-M^{N-i-1}+1,k} &  &  \\ 
&  & \ddots &  \\ 
&  &  & u_{jM^{N-i}-M^{N-i-1}+1,k} \\ 
&  &  & 
\end{array}
\right] _{M^{N-1}\times M^{N-i-1}}  \eqnum{XI}
\end{equation}
where only the elements $((j-1)M^{N-i}+(l-1)M^{N-i-1}+k,k)$ $(l=1,2,\cdots
,M;k=1,2,\cdots ,M)$ are nonzero. The nonzero elements of the first columns
in the different sub-blocks are equal to each other for the same index $l$.
We can find there are some self-similar property on the position of the
nonzero elements in unitary matrices $U_i$ and $U_{i+1}.$ In this special
case, our main task is to find the number of the least linear independent
columns in the unitary matrices $U_i(i=2,3,\cdots ,N-1).$ Using the same
discussion in the bi-party case, we find that there are at most $%
(M-1)M^{N-2} $ columns in the unitary matrices $U_i(i=2,3,\cdots ,N-1)$ are
linear dependent on the columns in the unitary $U_1,$ and at most $pM^{N-2}$
columns in matrix $U_{N-1}$ are linear dependent on the columns in the
matrix $U_1$. This can be reached by letting the elements $%
((k-1)M^{N-2}+j,j)(j=1,2,\cdots ,M^{N-2})$ be equal to 1 in the $k$th block
of the unitary matrix $U_1$. We use the same method to consider the columns
remaining in each unitary until the remaining columns only in the unitary $%
U_{N-1}$. Then we sum up all of the number of the linear independent columns
in each unitary matrix, we get 
\begin{equation}
(M-1)M^{N-2}+(M-1)M^{N-3}+\cdots +(M-1)M+p=M^{N-1}-M+p.  \eqnum{7}
\end{equation}
Since the $N$-partite state is a pure state, the nonzero eigenvalues of the
density matrix $\rho _N$ is the same as the nonzero eigenvalues of the
density matrix $\rho _{12\cdots N-1}$ as the Schmidt theorem. So the density
matrix $\rho _{12\cdots N-1}$ has at most $M$ nonzero eigenvalues and the
first $M^{N-1}-M$ eigenvalues are zeroes. The relations between the
eigenvalues of the one-party density matrices $\rho _i(i=1,2,\cdots ,N-1)$
and the global density matrix $\rho _{12\cdots N-1}$ become the relations
between the eigenvalues of the one-party reduced density matrices $\rho
_i(i=1,2,\cdots ,N-1)$ and $\rho _N.$ So the relations between the
eigenvalues of the one party density matrix $\rho _i(i=1,2,\cdots ,N)$ are 
\[
\sum_{j=1}^{N-2}\sum_{i=1}^{M-1}\lambda _i^{\uparrow
}(j)+\sum_{i=1}^p\lambda _i^{\uparrow }(N-1)\geq \sum_{i=1}^p\lambda
_i^{\uparrow }(N). 
\]
This is the end of the proof of the theorem 6. QED

From this proof, we can find that the nontrivial relations between the
one-party reduced density matrices of the multi-partite pure state in this
situation must include at least $(M-1)(N-2)+1$ eigenvalues (the same
eigenvalues are calculated repeatedly) at the left of the inequality.
Further more, using our method we can find almost all of the linear
relations between the eigenvalues of the one-party reduced density matrixes,
only need carefully consider the number of the orthogonal vectors needed to
express all of the columns. We conjecture that the necessary and sufficient
conditions of the compatibility problem between the one-party density
matrices and a pure state can be expressed by the linear relations between
these eigenvalues of the local one-party density matrix, that is, the
necessary and sufficient conditions can form a polytope in the eigenvalues
space. If this conjecture is true, we can find the necessary and sufficient
conditions for the compatibility problem by our method. The necessary and
sufficient conditions of the compatibility between the single qutrit density
matrices and a pure state in $C^3\otimes C^3\otimes C^3$ is 
\begin{eqnarray}
\lambda _1^{\uparrow }(A)+\lambda _2^{\uparrow }(A) &\leq &\lambda
_1^{\uparrow }(B)+\lambda _2^{\uparrow }(B)+\lambda _1^{\uparrow
}(C)+\lambda _2^{\uparrow }(C)  \nonumber \\
\lambda _1^{\uparrow }(A)+\lambda _3^{\uparrow }(A) &\leq &\lambda
_1^{\uparrow }(B)+\lambda _2^{\uparrow }(B)+\lambda _1^{\uparrow
}(C)+\lambda _3^{\uparrow }(C)  \nonumber \\
\lambda _2^{\uparrow }(A)+\lambda _3^{\uparrow }(A) &\leq &\lambda
_1^{\uparrow }(B)+\lambda _2^{\uparrow }(B)+\lambda _2^{\uparrow
}(C)+\lambda _3^{\uparrow }(C)  \nonumber \\
\lambda _1^{\uparrow }(A)+2\lambda _2^{\uparrow }(A) &\leq &\lambda
_1^{\uparrow }(B)+2\lambda _2^{\uparrow }(B)+\lambda _1^{\uparrow
}(C)+2\lambda _2^{\uparrow }(C)  \eqnum{8} \\
2\lambda _1^{\uparrow }(A)+\lambda _2^{\uparrow }(A) &\leq &\lambda
_1^{\uparrow }(B)+2\lambda _2^{\uparrow }(B)+2\lambda _1^{\uparrow
}(C)+\lambda _2^{\uparrow }(C)  \nonumber \\
2\lambda _2^{\uparrow }(A)+\lambda _3^{\uparrow }(A) &\leq &\lambda
_1^{\uparrow }(B)+2\lambda _2^{\uparrow }(B)+2\lambda _2^{\uparrow
}(C)+\lambda _3^{\uparrow }(C)  \nonumber \\
2\lambda _2^{\uparrow }(A)+\lambda _3^{\uparrow }(A) &\leq &2\lambda
_1^{\uparrow }(B)+\lambda _2^{\uparrow }(B)+\lambda _2^{\uparrow
}(C)+2\lambda _3^{\uparrow }(C)  \nonumber
\end{eqnarray}
and the conditions permutation $A,B$ and $C$. These conditions which are
found by Higuchi\cite{sudbery} can be found in our method more conveniently.
But it is very difficult to prove that this conditions are sufficient. This
necessary and sufficient conditions are linear and this support our
conjecture. When the number of the particle increasing, the simplexes of
this polytope increase rapidly, the proof used by Higuchi in the three
qutrit case is not convenient. We must need another method to proof the
convex property of this set.

\section{Summary}

In this paper we use a theorem of the analysis matrix to give a simple
method to find the relations between the reduced density matrix and the
multi-partite density matrix. We find hat the majorization relations are the
universal relations between the eigenvalues of reduced density matrices and
the multi-partite density matrix. We also give some relations of the
eigenvalues between the different reduced density matrixes. All of the
relations received in this paper can be viewed as the necessary conditions
of the problem whether the reduced density matrix is compatible with a
multi-partite density matrix. What is the necessary and sufficient
conditions for the compatibility problem between a arbitrary set of density
matrices and a multi-partite is far from completely solved, even the special
problem whether a set of one-party reduced density matrixes is compatible
with a pure multi-partite state is very difficult. But the method used in
this paper give us a possible way to solve these problems, especially for
the compatibility problem of the pure state. In this paper we only analyzed
the number of the linear independent vectors in the N-partite case, if we
analyze more carefully such as the coefficients of the linear combination,
we may find more relations. This is one of our further work. On the other
hand, if we add some symmetry such as the translation invariant on the
particles, we can get more constrains on the eigenvalues of the density
matrices, then we may find the necessary and sufficient conditions of the
compatibility problem. This symmetry is very useful in condensed matter and
the quantum phase transition. This problem will also be investigate in the
future.

The author would like to thank for invaluable discussion with L. M. Duan.
This work was funded by the National Fundamental Research Program
(2001CB309300), the Innovation Funds from Chinese Academy of Sciences (CAS),
the outstanding Ph. D thesis award (L.M.D) and the CAS's talented scientist
award (L.M.D.).

\end{document}